# Digital Image Mechanical Identification (DIMI)

Stéphane Roux and François Hild[*]

*LMT-Cachan*

*ENS de Cachan / CNRS UMR-8535 / Université Paris 6 / UniverSud Paris*

*61 avenue du Président Wilson*

*F-94235 Cachan Cedex, France*

---

[*] Corresponding author. Email: hild@lmt.ens-cachan.fr, Fax: +33 1 47 40 22 40.

# Digital Image Mechanical Identification (DIMI)

by

Stéphane Roux and François Hild

**Abstract**

A continuous pathway from digital images acquired during a mechanical test to quantitative identification of a constitutive law is presented herein based on displacement field analysis. From images, displacement fields are directly estimated within a finite element framework. From the latter, the application of the equilibrium gap method provides the means for rigidity field evaluation. In the present case, a reconditioned formulation is proposed for a better stability. Last, postulating a specific form of a damage law, a linear system is formed that gives a direct access to the (non-linear) damage growth law in one step. The two last procedures are presented, validated on an artificial case, and applied to the case of a biaxial tension of a composite sample driven up to failure. A quantitative estimate of the quality of the determination is proposed, and in the last application, it is shown that no more than 7% of the displacement field fluctuations are not accounted for by the determined damage law.

# Introduction

New advances in Solid Mechanics are linked to close and adequate interactions between experiments,[1] modeling and simulations.[2] During the last 50 years, this scientific field, among others, has experienced a complete revolution with the generalization of computers and the associated numerical techniques. In particular, the finite element method[3,4] has become a classical means used in various industries as a design tool, and is classically taught at graduate and even undergraduate levels. This progress opens the possibility to design reliable and sustainable structures *up to failure*. However, it first requires that numerical tools are accurate and validated, and that the used models are able to capture the most meaningful physical aspects of the material behavior. The validation steps call for intimate comparisons with relevant and well-calibrated experiments. The latter themselves are bound to evolve because of the need for renewed and more frequent interactions with numerical simulations. In their common future and close interactions probably lies the mastering of virtual design and / or virtual testing concept.

In experimental mechanics, the landscape is also evolving rapidly. Until recently, many design procedures relied on numerous experiments from coupons, to parts and even scale one structures (e.g., in aeronautical or automotive applications). The latter are very expensive and there is a clear industrial demand for their reduction or even ideally their substitution in the near future. The hope is to make virtual testing a reality. One of the needs is then related to the prediction of damage and fracture as a multiscale phenomenon. In particular, the scale effects are to be fully understood and quantitatively modeled to avoid the most expensive experiments. This task itself is one of the challenges facing computational and experimental mechanics for today and tomorrow.

In order to facilitate the interactions between experiments and simulation, a common language is desirable. One of the basic information used in numerical simulation is given by kinematic fields. The recent developments and generalization of *full-field* measurement techniques (i.e., photomechanics[5-9]) will definitely contribute to bridging the gap between computational and experimental mechanics. In the last years, digital image correlation techniques have been adapted to deal with kinematic fields that fulfill the same hypotheses as those used in finite element techniques.[10,11] More recently, even eXtended Digital Image Correlation algorithms[12] are developed in parallel to eXtended Finite Element Methods.[13] The latter technique allows for a very efficient modeling of discontinuities (e.g., fracture or localized shear bands) within the finite element framework and without any need for remeshing. Extensions of the same strategy for digital image correlations provide an efficient way of benefiting both from the accuracy of continuous displacement fields decomposed onto finite element shape functions over a regular square mesh, and still account for potential discontinuities (or rapid variations) of the displacement field.[14] All these tools combined together will allow for better calibration and tuning of constitutive laws and failure models. The present contribution aims at showing how mere images of a sample surface can be used to identify a damage model.

Different identification technbiques have been proposed to identify, say, elastic properties. Most of them differ from the type of norm used to minimize the gap between measured displacement fields (or derived strain fields) and predicted counter-parts.[15] In the following, a non-linear damage law is sought. The use of full-field data to feed an identification procedure for a damage law has already been proposed in the past. A first attempt was proposed by Claire *et al.*[16] using the same experiment as the one chosen herein. In this case, the relation between damage and an equivalent scalar strain was fitted to the collection of all element damage estimated from the Equilibrium Gap Method[17,18]. Another approach was proposed by Chalal *et al.*[19,20] based on the virtual fields method. A uniaxial composite loaded in shear was described by a linear increase of damage with the strain, leading to a parabolic stress-strain curve in Ref.[19], (a higher polynomial order without threshold strain was used in Ref.[20]). These different works open the way to identification techniques based on full field measurements.

In the present study, first the fundamentals of digital image correlation and the equilibrium gap method are presented, although a more detailed presentation can be found elsewhere (see e.g. Besnard *et al.*[11]). The strategy to identify a constitutive law, based on the previous tools and concepts, is presented and discussed for an isotropic damage law. In particular, a reconditioning strategy is proposed to provide a better robustness of the formulation. The resulting scheme is then tested on an artificial test case (computed from a known law), and whose displacement field is used blindly to mimic the result of a DIC procedure. After a successful validation with or without noisy data, the same procedure is applied to an experimental case. It consists in a biaxial tension of a cross shaped specimen of composite material up to failure. Different forms of damage laws are tested and shown to explain about 95% of the measured displacement fluctuations. A summary and prospective view is proposed in the last section.

## Image Correlation in a finite element formalism

The aim of the present section is to recall the main features associated with "finite element" digital image correlation. Two strategies are currently investigated, namely, the introduction of finite element kinematics in the correlation product[10] or in the optical flow conservation principle.[11] The latter is used herein since it provides the variational principle to solve the measurement problem.[21] As such, it is closer, from a conceptual point, to the finite element method in Structural Mechanics, which is based upon the virtual work principle (or the minimum total potential energy principle when applied to elasticity).

In the following, each displacement component is assumed to be decomposed as a linear combination of scalar shape functions $N_n(x)$ so that in an element $e$, the displacement vector $\mathbf{u}_e(\mathbf{x})$ reads

$$\mathbf{u}_e(\mathbf{x}) = \sum_n \sum_\alpha a_{\alpha n} N_n(\mathbf{x}) \mathbf{e}_\alpha \qquad (1)$$

where $n$ labels the different shape functions, $\alpha$ the space directions (e.g., 2 for plane problems as discussed herein), and $a_{\alpha n}$ denotes the unknown degrees of freedom to be determined by a suitable pattern matching algorithm. In the following, Q4 elements are considered with P1 interpolations. Consequently, each element $\Omega_e$ is mapped onto the unit square $[-1, 1] \times [-1, 1]$ where the four (i.e., $n = 4$) basic functions are $\frac{1}{4}(1 \pm x)(1 \pm y)$ in a local frame $(x, y)$. With this kinematic hypothesis and the use of two pictures $f$ and $g$ of the reference and deformed state of the observed surface, the unknowns $a_{\alpha n}$ are sought. The variational formulation consists in minimizing the following quadratic functional[11] (obtained from linearization)

$$\Phi_{lin}^2(a_{\alpha n}) = \sum_e \int_{\Omega_e} \left[ \sum_n \sum_\alpha a_{\alpha n} N_n(\mathbf{x}) \nabla f(\mathbf{x}) . \mathbf{e}_\alpha + f(\mathbf{x}) - g(\mathbf{x}) \right]^2 d\mathbf{x} \qquad (2)$$

with respect to $a_{\alpha n}$. For the convenience of numerical implementation, a single scalar label is introduced for the double indices as $j(\alpha,n)$. Since $\Phi_{lin}^2$ is a quadratic form of $a_{\alpha n}$, its minimization yields a linear system

$$[M]\{a\} = \{b\} \qquad (3)$$

in which $[M]$ is a matrix obtained as the assembly of elementary matrices $\lfloor M^e \rfloor$ whose components read

$$M^e_{j(\alpha,n)k(\beta,m)} = \int_{\Omega_e} \left(\partial_\alpha f(\mathbf{x})N_n(\mathbf{x})\partial_\beta f(\mathbf{x})N_m(\mathbf{x})\right)d\mathbf{x} \quad (4)$$

and the vector $\{b\}$ corresponds to the assembly of elementary vectors $\{b^e\}$ such that

$$b^e_{j(\alpha,n)} = \int_{\Omega_e} [g(\mathbf{x})-f(\mathbf{x})]\partial_\alpha f(\mathbf{x})N_n(\mathbf{x})d\mathbf{x} \quad (5)$$

with $\partial_\alpha f = \nabla f . \mathbf{e}_\alpha$. Additional details concerning multi-scale features as well as sub-pixel evaluations are given in Ref. [11] The main output of the correlation algorithm is therefore a Q4P1 displacement field that is post-processed to identify parameters of a constitutive equation, with the potential of exploiting the same discretization, i.e. without spurious loss of information based on independent meshes, since as a field is transported from one mesh to another one, its property of being the solution of a given mechanical finite-element problem, or a DIC solution will not be exactly preserved, henceforth leading to un-necessary inaccuracies. The foundation for the identification is to be found in the following Equilibrium Gap Method.

## Principle of the Equilibrium Gap Method

Among some of the identification techniques making use of full-field measurements,[15] it is proposed to extend the equilibrium gap method to identify non-linear constitutive laws. The equilibrium gap method[17] was introduced to analyze heterogeneous elastic media and identify damage fields and constitutive laws.[16] The displacement field is assumed to be known (e.g., measured by image correlation), while the elastic properties or damage field are to be determined. If only kinematic information is available, no stress scale is given and thus all constitutive parameters are determined up to a stress scaling factor.

It is assumed that the behavior of the material is such that any *incremental* changes are mapped onto an elastic (although heterogeneous) problem, an assumption that encompasses damage and plasticity. For the sake of simplicity, in the sequel, a simple scalar form is assumed, namely, the Poisson's ratio will be considered as constant. Thus the local constitutive equation is expressed as

$$\boldsymbol{\sigma}(\mathbf{x}) = (1 - D(\mathbf{x}))\mathbf{C}\boldsymbol{\varepsilon}(\mathbf{x}) \tag{6}$$

where $\boldsymbol{\sigma}$ and $\boldsymbol{\varepsilon}$ denote respectively the stress and strain tensors, $\mathbf{C}$ the Hooke's tensor, and $D$ the scalar damage field. In the following, the notation, $d(\mathbf{x}) = 1 - D(\mathbf{x})$, is introduced for the *stiffness contrast*. The stress field is to be balanced, and thus, in the absence of body forces, it obeys

$$\mathbf{div}(\boldsymbol{\sigma}) = \mathbf{0} \tag{7}$$

When resorting to a finite element formulation of the problem, a weak form of the above equations is classically written in terms of the discretized variables, i.e., the nodal displacement vector, $\{u\}$, the standard stiffness matrix, $[K]$, the nodal force vector (resulting from the boundary conditions), $\{f\}$, and the element contrast vector $\{d\}$

$$[K(\{d\})]\{u\} = \{f\} \tag{8}$$

For nodes away from boundaries, the nodal forces are equal to 0, thereby enforcing the balance condition between (here four) elements. Let us note that because of the simple choice of the constitutive law, each elementary stiffness matrix relative to a single element is simply linear in $d$, and involves the elementary stiffness matrix of a homogeneous and undamaged solid $[k^0]$

$$K_{ij} = \sum_e d_e k_{ij}^0 \tag{9}$$

where the sum over $e$ runs over elements sharing the nodes $i$ and $j$. The standard use of the finite element method to solve such an elastic problem is to assemble the stiffness matrix, $[K]$, and search for the displacement vector $\{u\}$. The problem one would like to solve here is different, namely, $\{u\}$ is assumed to be known while $\{d\}$ is to be determined. However, the writing of the basic equation (3) is strictly identical. The above equation, for the problem at hand is written as

$$L_{ie}d_e = 0 \tag{10}$$

where $i$ is an internal node index, and $e$ an element index. The matrix $[L]$ is rectangular and depends (linearly) on the measured displacement $\sum_e L_{ie} = K_{ij}u_j$. As formulated, the problem is over-determined since one has two in two dimensions (or three in three dimensions) such balance conditions at each node, while the unknown is one scalar per element. Thus it is generally impossible to fulfill all the above conditions, and one resorts to a weak form by minimizing the "equilibrium gap" vector norm, $E_g$

$$E_g = \sum_{Internal\ nodes} \|[L]\{d\}\|^2 \tag{11}$$

and hence solves the following linear system

$$[L]^t[L]\{d\} = \{0\} \tag{12}$$

As such this system is *not invertible* because as above mentioned, the contrast is only defined up to a constant scale factor. Thus this system is supplemented by an additional condition, which is most conveniently expressed as linear in $\{d\}$. For instance, one may choose one reference element for which $d_e$ is defined to be unity,[18] or alternatively impose that the arithmetic mean of the contrast is equal to unity. Any such choice is valid, and leads to an *invertible* linear system

$$[\bar{L}]\{\bar{d}\} = \{\bar{h}\} \tag{13}$$

The continuum analog of Equation (11) is

$$E_g = \int_D \|\mathbf{div}[d(\mathbf{x})\mathbf{C}\boldsymbol{\varepsilon}(\mathbf{x})]\|^2\, d\mathbf{x} \tag{14}$$

Let us note that the integrand involves a second-order differential operator acting on the experimentally determined displacement field, which signals an intrinsic fragility of the above formulation with respect to noise.

One way to mend this a priori sensitivity is to use a regularized form of the damage field.[22] That is, the damage field may be defined on a mesh that is coarser than that used for the displacement field. This hypothesis provides a spatial regularization, in the sense that it penalizes rapid variations. However, such a smooth variation may be a rather poor approximation for real damage problems that typically show a tendency to localize in high strain and damage regions, hence with steep gradients.

A second strategy, applied in the sequel and presented in the next section, is to use the assumption of a homogeneous but non-linear behavior. Therefore, the local damage variable will be assumed to be a smooth function of the local strain, rather than the spatial coordinate. This may allow for localization, and yet reduces drastically the unknowns of the problem, hence providing the sought regularization.

However, this basic formulation is still based on a second order differential operator, and although the number of unknowns in the problem is drastically reduced, noise sensitivity may be feared. A reconditioning technique is proposed to bring back the operator acting on the fields to a differentiation degree equal to 0, as explained in the second next section.

## Identification of a constitutive law

Up to now, the damage field was introduced to account for a heterogeneous stiffness, but it did not result from a damage constitutive law. This is now introduced. The analysis performed herein is based upon an isotropic damage description.[23] A classical continuum thermodynamics setting is used.[24] Under isothermal conditions, the material state is described by the infinitesimal strain tensor $\varepsilon$ and the damage variable $D$ (with its usual bounds, namely, $D = 0$ for a virgin material and $D = 1$ for a fully damaged state) so that the state potential $\psi$ (i.e., Helmholtz free energy density) reads

$$\psi = \frac{1}{2}(1-D)\boldsymbol{\varepsilon} : \mathbf{C} : \boldsymbol{\varepsilon}, \qquad (15)$$

where ':' denotes the contraction with respect to two indices. In Equation (15), only a recoverable part of the state potential is considered. Consequently, it is assumed that no residual stresses are present, created or relaxed within the material during the whole load history. The associated forces to the state variables are respectively defined by

$$\boldsymbol{\sigma} = \frac{\partial \psi}{\partial \boldsymbol{\varepsilon}} = \mathbf{C}(1-D)\boldsymbol{\varepsilon} \quad \text{and} \quad Y = -\frac{\partial \psi}{\partial D} = \frac{1}{2}\boldsymbol{\varepsilon} : \mathbf{C}\boldsymbol{\varepsilon}, \qquad (16)$$

where $Y$ is the energy release rate density.[25] The thermodynamic force $Y$ under plane stress assumption is computed from the in-plane strain field in a non-dimensional way

$$\frac{Y}{E_0} = \frac{\varepsilon_{11}^2 + 2\nu_0 \varepsilon_{11}\varepsilon_{22} + \varepsilon_{22}^2}{2(1-\nu_0^2)} + \frac{\varepsilon_{12}^2}{1+\nu_0}, \qquad (17)$$

where the directions 1 and 2 are associated to an in-plane frame, $E_0$ the Young's modulus of the virgin material and $\nu_0$ the corresponding Poisson's ratio.

Clausius-Duhem inequality, in the present case, reduces to

$$Y\dot{D} \geq 0, \qquad (18)$$

where a dotted variable corresponds to its first time-derivative. Since the energy release rate density $Y$ is a positive function, the damage growth is such that

$$\dot{D} \geq 0. \qquad (19)$$

Within the framework of generalized standard materials[26] and for a time-independent behavior, damage growth is written as[27]

$$\dot{D} = \dot{\delta}\frac{\partial f}{\partial Y}, \qquad (20)$$

where the damage multiplier $\dot{\delta}$ satisfies the Kuhn-Tucker conditions, and $f$ is the loading function. For concrete-like materials, another choice is given by Mazars' law[28]

$$D(t) = H\left[\max_{0 \leq \tau \leq t} \varepsilon_+(\tau)\right] \quad \text{with} \quad \varepsilon_+ = \left(\langle\varepsilon_1\rangle^2 + \langle\varepsilon_2\rangle^2 + \langle\varepsilon_3\rangle^2\right)^{1/2} \qquad (21)$$

where $H$ is a monotonically increasing function to be identified, $\varepsilon_i$ the eigen strains ($i = 1$ and 2 are in the observation plane, and 3 is normal to it), and $\langle\bullet\rangle$ the Macauley brackets (i.e., positive part of the inner argument). From the above presentation, an equivalent strain, $\varepsilon^{eq}$, is the driving force of the damage (or contrast) variable. In the present formalism, the equivalent strain is evaluated from the experimentally measured displacement field. Two different forms are considered under a plane stress assumption, here written in terms of principal strains. First, what will be referred to as Marigo's form (equivalent to Equation (17))

$$\varepsilon^{eq} = \left[\frac{\langle\varepsilon_1\rangle^2 + 2\nu_0\langle\varepsilon_1\rangle\langle\varepsilon_2\rangle + \langle\varepsilon_2\rangle^2}{2(1-\nu_0^2)}\right]^{1/2} \qquad (22.a)$$

and, second, Mazars' form

$$\varepsilon^{eq} = \left[\langle\varepsilon_1\rangle^m + \langle\varepsilon_2\rangle^m + \frac{\nu_0}{(1-\nu_0)}\langle-\varepsilon_1-\varepsilon_2\rangle^m\right]^{1/m} \qquad (22.b)$$

where $m$ is a positive number, whose influence will be investigated.

The multiaxial strain state (under plane stress condition) is written as

$$\begin{cases} \varepsilon_1 = A\cos(\theta) \\ \varepsilon_2 = A\sin(\theta) \end{cases} \qquad (23)$$

so that $\varepsilon^{eq}(\theta) = A\Xi(\theta)$. A constant damage level corresponds to a constant equivalent strain, and hence the strain magnitude $A$ is proportional to $1/\Xi(\theta)$. To illustrate the difference between these different criteria, Figure 1 shows a polar plot of $1/\Xi(\theta)$. As will be shown further down, uniaxial strain states (within the plane) play a dominant role and hence Marigo's form is rescaled by a factor $\sqrt{2(1-v^2)}$ to match Mazars' form for $\theta = 0$. The criteria mostly differ in the bi-compression direction (Figure 1).

In both cases the unknown function $H$ depends on the maximum equivalent strain $\hat{\varepsilon}^{eq}$

$$D = H\left(\hat{\varepsilon}^{eq}; c_p\right) \quad \text{with} \quad \hat{\varepsilon}^{eq}(t) = \max_{0 \leq \tau \leq t} \varepsilon^{eq}(\tau) \tag{24}$$

where $c_p$ are parameters to be determined. A convenient form of this relationship is for instance

$$H\left(\hat{\varepsilon}^{eq}; c_p\right) = \sum_p c_p \varphi_p\left(\hat{\varepsilon}^{eq}\right) \tag{25}$$

where $\varphi_p$ are chosen functions. In the following, we selected $\varphi_p(\hat{\varepsilon}^{eq}) = 1 - \exp(-\hat{\varepsilon}^{eq}/\varepsilon_p)$, for a series of characteristic strains $\varepsilon_p$. Note that the following form for $H$ is not restrictive as it may be read as a Laplace transform of $H$. Equation (25) is a constitutive law regularization. It enforces the strong condition that two regions where the equivalent strain is the same will experience an identical damage. It is impossible to fulfill exactly the above condition, and thus we resort to a least square minimization, namely, the parameters $c_p$ are searched for in order to minimize the equilibrium gap

$$E_g(c_p) = \sum_i \left( \sum_e L_{ie}\left[1 - H(\hat{\varepsilon}_e^{eq}; c_p)\right] \right)^2 \tag{26}$$

Using Equation (25), the minimization leads to

$$\sum_p \sum_{e,f} \left[ \sum_i L_{ie} L_{if} \right] \varphi_p(\hat{\varepsilon}_e^{eq}) \varphi_q(\hat{\varepsilon}_f^{eq}) c_p = \sum_{e,f} \left[ \sum_i L_{ie} L_{if} \right] \varphi_q(\hat{\varepsilon}_f^{eq}) \tag{27}$$

One does not need to resort to an additional condition to set the contrast scale if the condition $\varphi_p(0) = 0$ is introduced. However, if some chosen test functions do not vary significantly over the range of equivalent strains covered by the displacement field data, their amplitude cannot be determined. In practice, the minimum (resp. maximum) characteristic strain has to be larger (resp. less) than the minimum (resp. maximum) values of equivalent strains.

Let us note that the algorithm becomes unstable with respect to the introduction of test functions with a very small $\varepsilon^p$. As this value becomes small as compared to the set of equivalent strain values explored in the measured displacement fields, any amplitude would correspond to an initial damage in the system that cannot be determined. Thus it is important to introduce progressively smaller values of $\varepsilon^p$ until an optimal identification quality (to be defined further down) is obtained.

Although the initial problem is highly non-linear, let us underline that the final formulation is a *linear* problem. The non-linearity of the equivalent strain with respect to the displacement is a local problem computed once for all from the experimentally determined displacement field. The non-linearity of damage growth with the equivalent strain is included in the choice of the $\varphi$ functions. The drawback for such a simplification is that there is no guarantee that the resulting damage law will not assume either negative (or larger than one) values. Therefore, a further step of projection of the amplitudes $c_p$ to positive values is introduced, imposing a "Kuhn-Tucker" condition, with a conjugated "force" $f_p$ to the amplitude $c_p$, such that

$$f_p = \frac{\partial E_g}{\partial c_p} \qquad (28)$$

and thus either $c_p = 0$, when $f_p \geq 0$, or $c_p \geq 0$, when $f_p = 0$. This sets the problem as a simplex one; it ensures Clausius-Duhem inequality (18) to hold, and thus thermodynamic consistency of the identified damage model. It is straightforward to implement additional constraints such as $\sum_p c_p = 1$ (for the present choice of test functions) to ensure that the damage reaches 1 for large strains, or inequalities $\sum_p c_p \leq 1$ so that the damage remains smaller than unity.

Lagrange multiplier technique is suited for this purpose. However, we did not implement such a procedure for the cases considered below.

## Reconditioning

Equation (26) deserves some additional comments. As earlier mentioned, the matrix $[L]$ is linear in the measured nodal displacement $[L] = [K]\{u_{meas}\}$, and involves the (homogeneous) elastic properties of the *undamaged* material. The sum $\sum_e L_{ie} = f_i$ is interpreted as a nodal resulting force. The elastic problem with an undamaged material, with Dirichlet boundary conditions (imposed displacement at the boundary), and known body forces is well-posed and is inverted to compute the displacement in the bulk, (or the nodal displacement in a finite element formulation)

$$\{u\} = [S]\{f\} \qquad (29)$$

where $[S] = [K]^{-1}$ is the inverse stiffness matrix (of the undamaged material).

Considering the poor conditioning expected from the fact that $[L]$ is a second order differential operator, it is natural to substitute to the initial proposition (26) the following form

$$\widetilde{E}_g(c_p) = \left\| \sum_j S_{ij} \sum_e L_{je} \left[1 - H(\hat{\varepsilon}_e^{eq}; c_p)\right] \right\|^2 \qquad (30)$$

This rewriting is further simplified by exploiting the property $[S][L] = \{u_{meas}\}$, so that

$$\widetilde{E}_g(c_p) = \left\| u_{imeas} - \sum_j S_{ij} \sum_e L_{je} H(\hat{\varepsilon}_e^{eq}; c_p) \right\|^2 \qquad (31)$$

This is the final form proposed herein. It is to be noted that the reconditioning has brought back the expression inside the norm as equivalent to a displacement, so that it cancels precisely the double differentiation that was suspected to be the weak point of the initial formulation. The unknowns are thus treated at the same level as the measured displacement field. One cannot further simplify the writing of Equation (31), since the damage field varies at the scale of elements. The introduction of the compliance matrix [S] may appear to be a prohibitive cost. However, it should be noted that [S] has not to be explicitly computed. Rather, one should solve an elastic and homogeneous problem (it is recalled that [S] refers to the undamaged material) with Dirichlet boundary conditions, for each choice $p$ of the basis functions, and for each loading stage. Moreover, only fictitious body forces vary with the different values $p$ while the stiffness matrix remains constant.

## Validation on a computed case

It is important to validate the proposed scheme. This is performed by resorting to an artificial case that is computed with a known constitutive law. The set of displacement fields is then used blindly as the "measured" one for identification purposes. The resulting law is then compared with the known one. A uniaxial compression of a square sample is considered under plain stress assumptions. To introduce some heterogeneity, the top and bottom sides are subjected to a uniform translation with no transverse displacements. The damage field is heterogeneous as the applied displacement amplitude increases (Figure 2). The constitutive law is chosen to be of Mazars' type, with an exponent $m = 2$, and $H(\varepsilon) = 1 - \exp(-\varepsilon / \varepsilon^0)$, with $\varepsilon^0 = 10^{-3}$. The Poisson's ratio is taken equal to 0.20. The domain is partitioned as a $64 \times 64$-element mesh, using Q4P1 finite elements, i.e., the *same* basis as that at the measurement stage. The displacement field is solved for 6 different stages of loading from the elastic regime to the onset of localization. The displacement field corresponding to the final step is shown in Figure 3.

In order to really test the algorithm itself, the actual damage law is not included in the set of trial functions $\varphi_p$. Four such functions were chosen $\varphi_p(\varepsilon) = 1 - \exp(-\varepsilon / \varepsilon^p)$, with $\varepsilon^p = 0.35 \times 10^{-3}$, $0.7 \times 10^{-3}$, $1.4 \times 10^{-3}$ and $2.8 \times 10^{-3}$ for $p = 1, 2, 3$ and $4$. The identified amplitudes were $c_p = 0.000$, $0.487$, $0.510$, and $0.003$, respectively. A comparison between the identified and prescribed damage laws is proposed in Figure 4a. It is worth noting that the maximum equivalent strain that is experienced in the set of different loading stages used in the analysis is $1.7 \times 10^{-3}$. The agreement is quite good over this range whereas a slight difference appears for larger strains not encountered in the analyzed field (see Figure 4b). Let us also note that the equality $\sum_p c_p = 1$ is precisely satisfied, although no such constraint was prescribed.

A more significant test is to consider the displacement field, $\mathbf{u}_{computed}$, which is computed using the identified law, and prescribing the measured displacement along the boundary. Note however, that the local damage required for this computation is evaluated based on the *measured* displacement, rather than the *computed* one. From the computed displacement field a relative error, $\rho$, is proposed. It is defined as the ratio of the standard deviation (denoted by $\chi(.)$) of the difference between the measured and computed displacement field, normalized by the standard deviation of the measured displacement field

$$\rho = \frac{\chi(\mathbf{u}_{computed} - \mathbf{u}_{measured})}{\chi(\mathbf{u}_{measured})} \qquad (32)$$

This relative error measures in a non-dimensional way the decorrelation degree of the measured displacement field by the identified damage law, and ideally reaches 0. In the present case, the highest relative error is 0.002 (reached for the final loading step). Figure 3 shows the measured and identified displacement fields along both directions. A very good agreement is observed as expected from the very low value of $\rho$.

We also tried to identify the damage growth law, by using a different exponent $m = 1$, instead of the value $m = 2$ used in the generation of the case study. The damage law cannot be directly compared since the definition of the equivalent strain is different. However, the quality of the displacement field obtained with the identified law was also very good, i.e., $\rho = 0.005$, almost as good as with the correct exponent. Thus, at least for this test case, the algorithm does not appear to be able to discriminate among exponent values. If a different criterion is used such as Marigo's form, the relative error increases up to a ten-times larger value (0.026).

In order to test the noise sensitivity of the procedure, the displacement field of the computed test case was corrupted by a Gaussian white noise of different amplitudes. At the last level of loading, the standard deviation of the displacement field without noise is $8.7 \times 10^{-4}$ and $1.5 \times 10^{-4}$ along the vertical (resp. horizontal) directions. For a low noise level, of the order of 1% or less of the displacement standard deviation (or $10^{-5}$ in the present case), the identified law is not affected at all by the presence of noise. Therefore a very high noise level of 10% of the maximum displacement range encountered in the load history was added to each component of the displacement. The Gaussian distributed noise, without correlation in space or time, of zero mean and given variance is added to each of the 6 stages of loadings. Let us underline that this level of noise is much larger than the level anticipated in the real experimental case studied in the following section. This level of noise induces a severe deterioration of the relative error that reaches 0.19 at the final loading stage. Let us however note that this relative error is estimated from the *noisy* displacement data.

Figure 5 shows the noise-corrupted displacement field at the final stage of loading, the computed one from the identified law and the difference between both fields. The identified displacement field is almost deprived from any short scale fluctuations apart from the boundary where the noisy displacements are prescribed. Conversely, the difference gathers most of the noise. The identification acts as a filter that is extremely efficient as the displacement field is far from being homogeneous. To further check this filtering property, the noise used for corrupting the displacement data is shown in Figure 6. Along the transverse direction, the residual difference is extremely close to the initial noise, not only in terms of order of magnitude, but also similar patterns are recognized in both maps. Along the (loading) *x*-direction, the residual is at most twice the noise level, and part of the constitutive law has been missed. This excellent noise robustness is for a large part due to the reconditioning proposed in the previous section.

## Application to an experimental case

A vinylester matrix reinforced by E-glass fibers is studied. A quasi-uniform distribution of orientations leads to an isotropic elastic behavior prior to matrix cracking and fiber breakage, which are the main damage mechanisms.[29] A cross-shaped specimen (Figure 7a) is loaded in a multiaxial testing machine. The experiment is performed in such a way that the forces applied along two perpendicular directions are identical. Their norm is denoted by *F*. Eleven different load levels are analyzed, namely, $F = 1$ kN to 11 kN with respect to the reference for no applied load. The failure load corresponded to $F = 11.1$ kN (Figure 7b). The element size to perform the correlation analysis was chosen to be equal to $16 \times 16$ pixels (i.e., an area of 0.5 mm$^2$). It corresponds to a compromise between the spatial resolution and the displacement uncertainty. The latter is evaluated a priori by performing artificial sub-pixel translations varying from 0 to 1 pixel with increments of 0.1 pixel.[11] The maximum value for the standard deviation was found to be of the order of $7 \times 10^{-3}$ pixel. For each load level, 3565 displacement measurements are obtained.

The Poisson's ratio has been determined in a preliminary study[29] from a strain gauge placed on a specimen for small load levels. Its value was estimated to be $\nu = 0.28$. Rather than using this value directly, it is proposed to estimate this parameter from the displacement field measured at the first load level ($F = 1$ kN). It is assumed that damage is negligible, and the EGM is applied by assuming a uniform and isotropic elastic behavior. In this case, Equation (8) still holds for $d = 1$ for all elements. The property one wishes to exploit is the affine dependence of the rigidity matrix, [*K*], on the Poisson's ratio, $\nu$, if the combination $F = E/(1-\nu^2)$ (where $E$ is the Young's modulus) is considered as a fixed elastic modulus. As already mentioned, the absence of stress scale forbids one to determine $F$ (or equivalently $E$). Let us therefore write the stiffness matrix under the form

$$[K] = F\left([K_1] + \nu[K_2]\right) \tag{29}$$

where both "rigidity" matrices $[K_1]$ and $[K_2]$ consists of pure numbers (known for any shape function). Equation (8) thus takes the form

$$\left([K_1] + \nu[K_2]\right)\{u_{meas}\} = \{S_1\} + \nu\{S_2\} \tag{30}$$

where $\{S_1\}$ and $\{S_2\}$ are vectors computed from the boundary displacements. Paralleling the previously exposed identification approach, the above equation is recast in the least squares sense. The problem is solved for different trial values of the Poisson's ratio, and using a parabolic fit of the lower values of $\rho$, the minimum is easily computed. This allows one to estimate directly the optimum value of the Poisson's ratio. Figure 8a shows the objective function $\rho$, its parabolic fit, and the best value of the Poisson's ratio is estimated to be $\nu = 0.31$ leading to a relative error $\rho$ equal to 11%. Note that this value $\nu$ is quite close to the value which was estimated based on strain gauge measurements ($\nu = 0.28$).

The difficulty to be underlined however for this procedure is that one has to assume that the behavior is elastic for the first load level considered. In this case, however, the displacements are of low magnitude and hence the noise level in the data is rather high. This drawback may be counterbalanced by the fact that the error map at the optimal Poisson's ratio is a direct output of the computation, which may show the quality of the homogeneity assumption. Figure 8b shows the residual error map for this first load level assuming no damage. From this figure, it appears that presumably some damage was already present in the specimen prior to any loading. Alternatively, some slight heterogeneity and anisotropy of the material could also be responsible for the pattern observed. Note however that the magnitude of the difference between measured and identified displacement is low, in spite of the fact that the measured displacement is small. Furthermore, a similar analysis carried over for the second load level (tension of 2 kN applied in both directions) leads to a slightly larger estimate of $\nu = 0.34$.

In terms of damage law, this test is first analyzed by using Mazars' equivalent strain. The Poisson's ratio is taken equal to the above determined value 0.31 (an earlier trial with a value of 0.28 did not change the following results). Two values of the exponent, $m = 1$ and 2, are tested. Different characteristic strains in the damage law test functions were considered $\varepsilon^p = (0.005, 0.01, 0.02, 0.04, 0.08)$ consistently with the range of $\varepsilon_{eq}$ encountered in the experiments (Figure 9a). For $m = 1$, the five amplitudes were $C_p = (0.831, 0, 0, 0, 0.161)$, resulting in a damage law shown in Figure 9b when the last four pictures are considered. The quality of the analysis appeared to be good, namely $\rho = 0.05, 0.05, 0.05$ and $0.07$. Figure 10 shows a comparison between the measured and predicted displacement fields for the last two load steps. The fact that the quality deteriorates in this last level is due to a well-developed crack on the top left part of the sample. The crack is both crudely accounted for by the scalar damage model, but also presumably badly captured by the image correlation algorithm which is designed for continuous displacement fields.

For an exponent $m = 2$, the estimated coefficients are close to the previous ones $C_p = (0.807, 0, 0, 0, 0.193)$ when using the same values of characteristic strains. The quality was observed to be better, although the sensitivity to this parameter appears to be rather low. In that case, one eigenvalue of the strain is overwhelmingly important, and dominates the other ones. The following relative errors were obtained: $\rho = 0.03, 0.03, 0.03$ and $0.05$ for the four last loading steps. Figure 11 shows the comparison between the measured and identified displacement fields for the penultimate loading step (10). These results in terms of spatial distribution are very close to those obtained for $m = 1$ (Figure 10b).

Last, Marigo's law is also tested. Because of the definition of the equivalent strains given in Equation (22), they are comparable in the experiment reported herein. Figure 12 shows the correspondence between both equivalent strains at the last step of loading. A simple proportionality fits the data quite well. Assuming that the largest strains are essentially uniaxial, the theoretical slope should be $1/\sqrt{2(1-\nu^2)} \cong 0.74$. In Figure 12, Marigo's equivalent strain was rescaled by this factor and a straight line corresponding to the identity between both strains is shown as a reference. Note that this observation is merely factual and only holds for the present case since bicontraction would discriminate among the chosen criteria (Figure 1). The characteristic strains were chosen accordingly with this observation to match the values used for Mazars' law $\varepsilon^p = (0.0067, 0.0135, 0.027, 0.054, 0.108)$. The identified law was $C_p = (0.871, 0, 0, 0, 0.125)$. The quality is observed to be quite comparable at all steps of loading, namely, $\rho = 0.03, 0.03, 0.03$ and $0.05$. This result is confirmed by the displacement maps shown in Figure 13, when compared to Figures 10b and 11.

The damage maps are very informative (Figure 14). Crack initiation on the left hand and top corner of the specimen is clearly depicted, as also observed in the picture of step 11 (Figure 7a) when magnified. However, much prior to this loading, damage concentration is already seen at step 9. For this load level, the visual inspection of the picture does not reveal any crack initiation nor damage localization. Damage concentrations at other corners are also observed and correlated rather well with the final cracking pattern (Figure 7b).

## Summary and perspectives

Digital image correlation now offers the possibility of directly evaluating displacement field decomposed onto a finite element basis from sets of experimental images. From the latter kinematic analysis, an algorithm has been presented which aims at identifying a full damage law. An artificial case for validation, and an experimental case for illustration purposes have been documented. Excellent levels of performance have been obtained in terms of the reproduction of the measured displacement field from the identified damage law. Different forms of damage laws were observed to provide similar qualities, thus making it difficult to discriminate among them. This may however be attributed to the particular test case at hand, where the different forms of equivalent strains were observed to be close to proportional to one another.

Let us emphasize a few points making this technique very attractive:

- It makes use of the entire displacement field, thus keeping a very intimate connection to reality.
- It does not require any further information than a limited set of images (five in the present case, to obtain four displacement fields to identify the damage growth law). It is thus extremely tolerant to any imperfection in the experimental set-up, which is very advantageous for complex multiaxial mechanical tests.
- The core of the algorithm is linear, inasmuch as all non-linearities are local.
- Any form of damage growth law is simply incorporated in the set of trial functions, and a quality evaluation of the identification is given, which allows one to modify the set of proposed forms at will up to satisfaction.
- The identification computation time is about a minute on a personal computer.

Let us finally stress that although only damage has been discussed in the present contribution, extensions to any incrementally elastic problem does not require much further development of the present formalism. By exploiting the fact that the incremental change may be phrased as a (linear) elastic but inhomogeneous problem, the equilibrium gap method (and its reconditioned version) can be readily used. The point to be imagined is a parameterization that captures the history dependence and the hardening law of plasticity, to be used to ascribe the tangent modulus to each element consistently so that the parameters are introduced in the EGM functional.

Moreover extensions to other optical techniques for measuring displacements could also be considered,[8] as they may equally benefit from a close connection with identification techniques. However, to avoid reprojection errors, it is desirable to perform the experimental determination of the displacement on the same basis as that of the finite element method, the underlying numerical tool used in the identification procedure. In the present case, the measured kinematics corresponds to the degrees of freedom associated with shape functions of four-noded square finite elements.

## Acknowledgments

This work is part of a project (PHOTOFIT) funded by the Agence Nationale de la Recherche. The authors wish to thank Prof. Sylvain Calloch with whom the experiment reported herein was performed.

# Figure captions

Fig. 1.  Polar plot of $1/\Xi(\theta)$ defined in the text after Equation (23) for the three criteria used in the present study. This plot shows the effect of different strain biaxialities on the damage growth law.

Fig. 2.  Three damage maps, $D(x)$, at three stages of loading for the computed test case. The final one (c) is at the onset of localization.

Fig. 3.  Reference (left) and identified (middle) displacement fields, and difference between them (right). The top (resp. bottom) line refers to the horizontal (resp. vertical) components. The criterion chosen is Mazars' one with an exponent 2. Note that the color scale is different for the difference and computed displacements.

Fig. 4.  (a) Comparison between the imposed (dashed line) and identified (solid line) damage laws versus equivalent strain. Note that the actual law is not included in the set of trial functions, and hence a perfect agreement cannot be reached. (b) Histogram of the equivalent strains encountered in the six different displacement fields of the computed test case. The maximum strain is shown in Figure 4a (vertical dashed line) to depict the validity domain of the damage law. In the latter, the identification is very good.

Fig. 5.  Corrupted (left) and identified (middle) displacement fields, and difference between them (right). The top (resp. bottom) line refers to the horizontal (resp. vertical) components. The criterion chosen is Mazars with an exponent 2. Note that the colorscale is difference for the difference and computed displacements. The uncorrupted case corresponds to Figure 3.

Fig. 6. Noise maps associated with the horizontal (resp. vertical) components of the displacement field. These maps are to be compared with the displacement differences shown in Figure 5(right).

Fig. 7. 11$^{th}$ and 12$^{th}$ images taken from the biaxial tension test on a composite sample.

Fig. 8. (a) Evolution of the relative error with the trial Poisson's ratio (data point) and parabolic fit (solid line). The minimum is reached for $\nu = 0.31$. (b) Comparison between measured and identified displacement fields at the first step of loading. The residual displacements are also shown.

Fig. 9. (a) Log-log plot of the histograms of strain encountered in the displacement fields of the last four loading steps. (b) Identified damage law using the three different forms of equivalent strain. Note that Marigo's strain has been scaled to be comparable to the other ones.

Fig. 10. Comparison between measured, identified displacement fields at the final step of loading (a) and the penultimate one (b). The difference between the two displacements is also shown. The form of the damage law is here assumed to be of Mazars type, with an exponent $m = 1$. The relative error $\rho$ is about 7% (a) and 5% (b).

Fig. 11. Comparison between measured and identified displacement fields at the penultimate step of loading. The form of the damage law is here assumed to be of Mazars type, with an exponent $m = 2$. The relative error $\rho$ is about 3%.

Fig. 12. Correspondence between scaled Marigo and Mazars ($m = 2$) equivalent strains for all elements for the final loading step. The data points are shown as dots, while the solid line is the identity.

Fig. 13. Comparison between measured and identified displacement fields at the penultimate step (10) of loading. The form of the damage law is here assumed to be of Marigo type. The relative error $\rho$ is about 3%.

Fig. 14. Maps of $D$ for the last three steps of loading (9, 10, 11) obtained with Marigo's equivalent strain. One clearly sees in the left-hand top corner the initiation of a major crack that will lead to failure of the sample. Moreover secondary crack formations are also distinguished close to the other corners (see Fig. 7b for a detailed comparison with the final failure pattern).

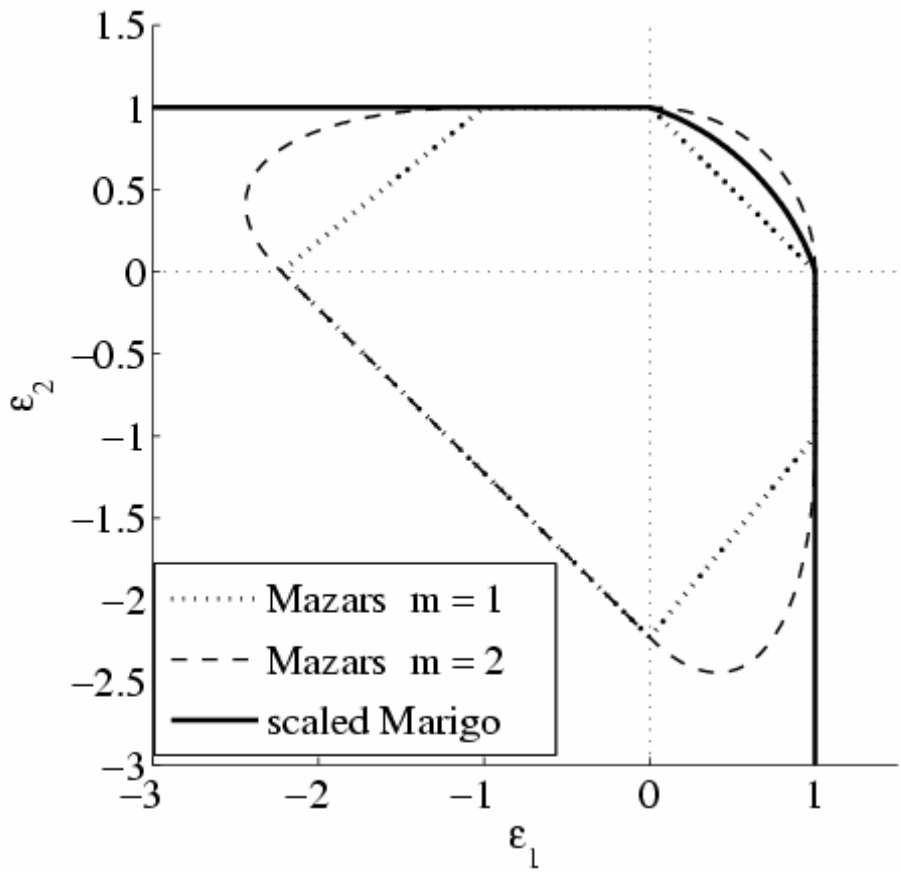

Fig. 1.

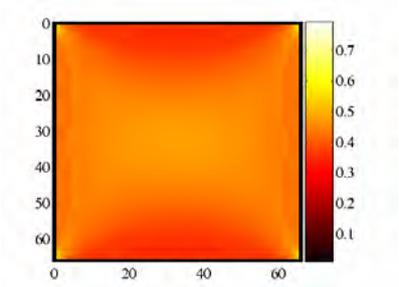 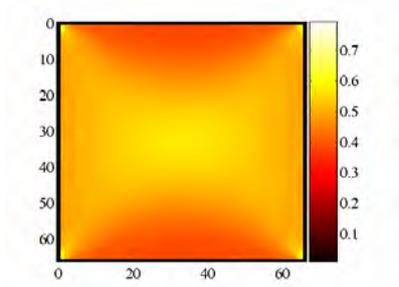 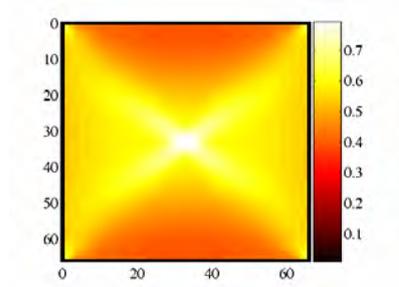

-a-          -b-          -c-

Fig. 2.

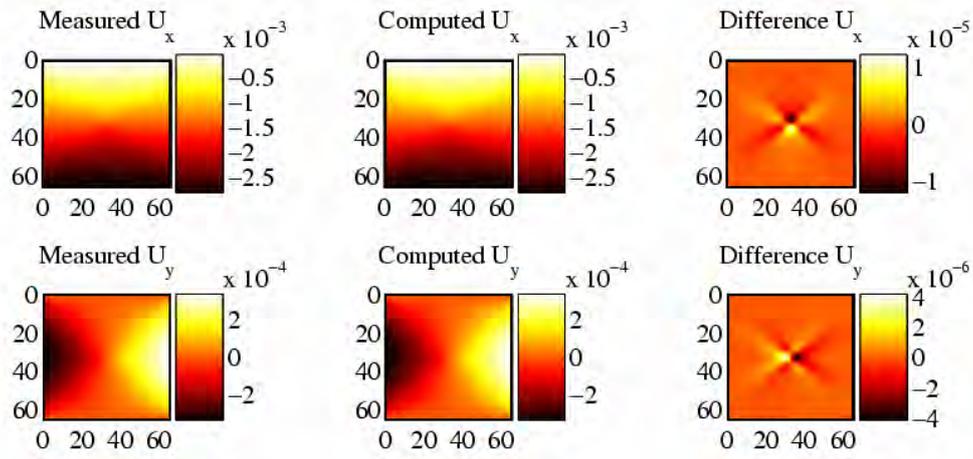

Fig. 3.

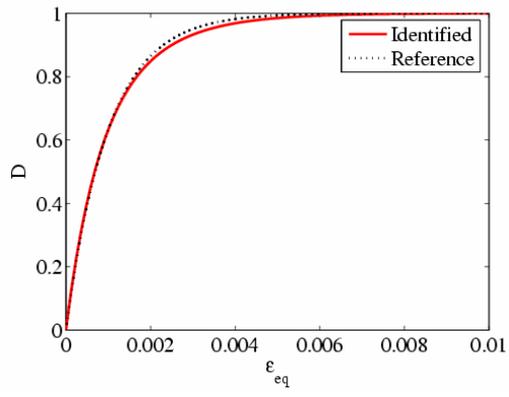 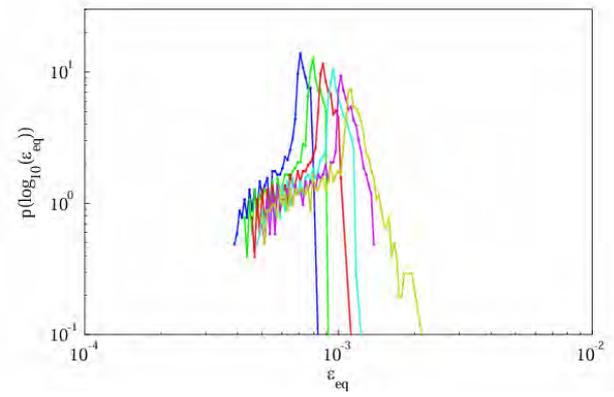

-a-              -b-

Fig. 4.

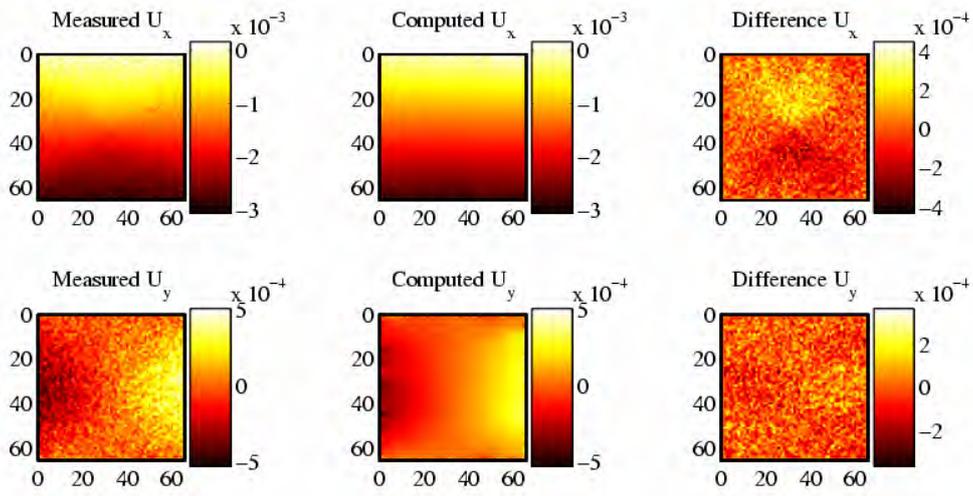

Fig. 5

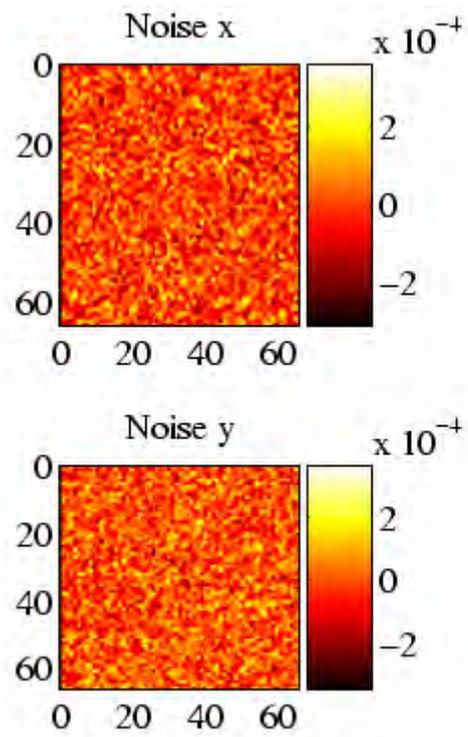

Fig. 6

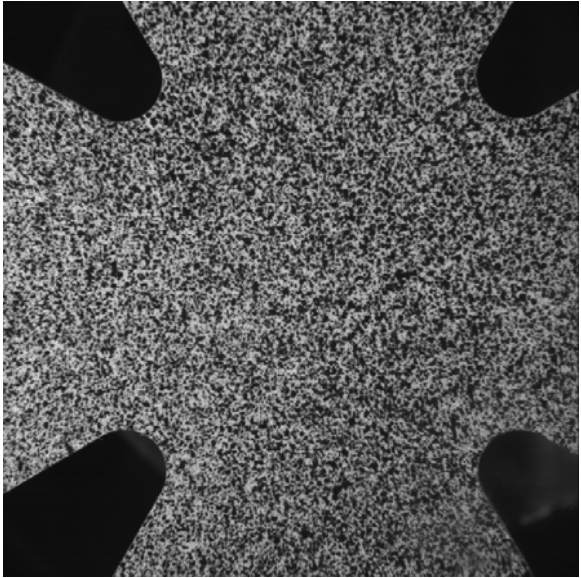 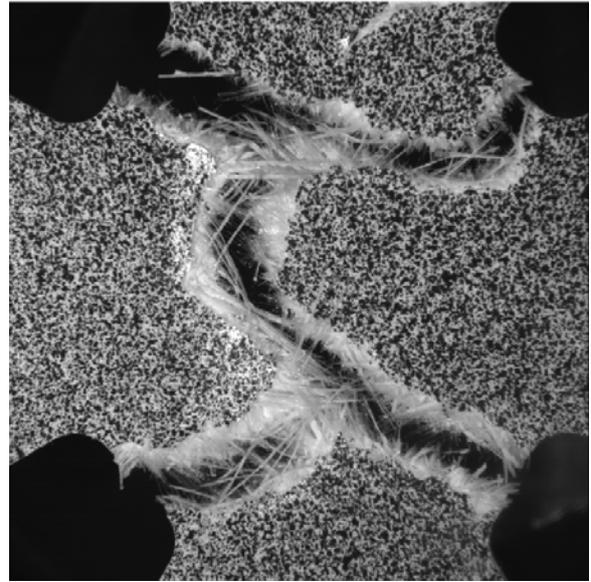

-a- -b-

Fig. 7

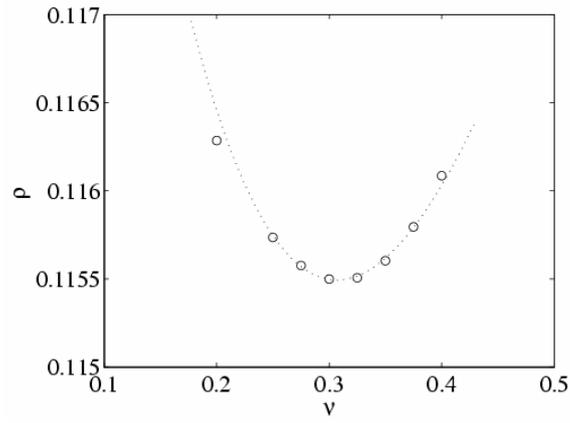

-a-

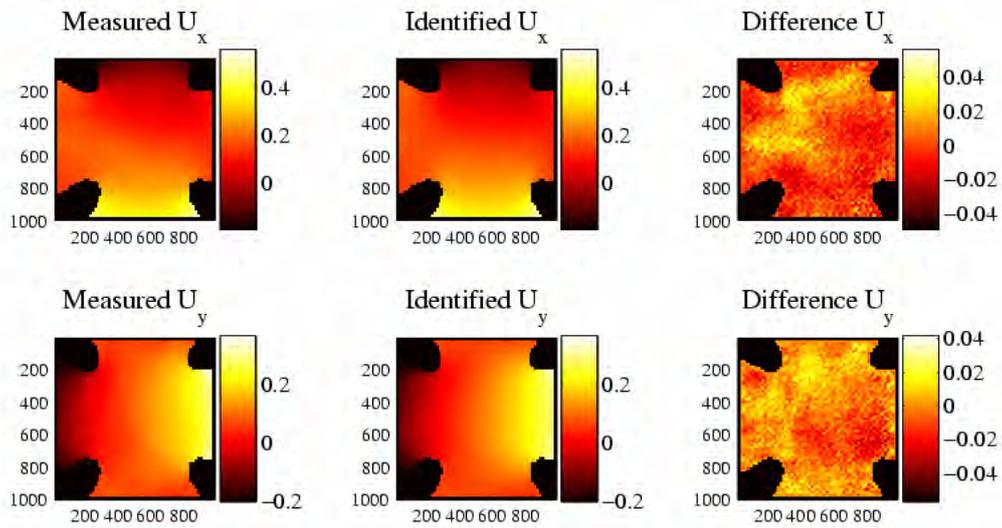

-b-

Fig. 8.

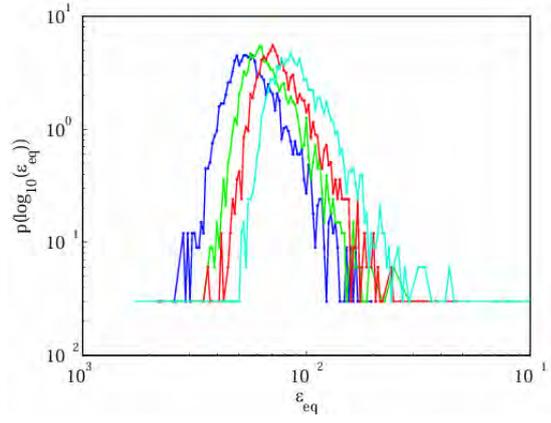 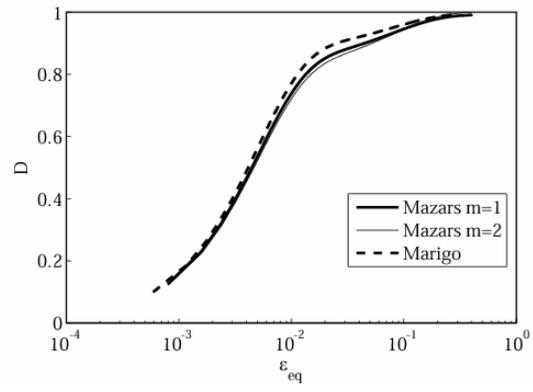

-a- -b-

Fig. 9.

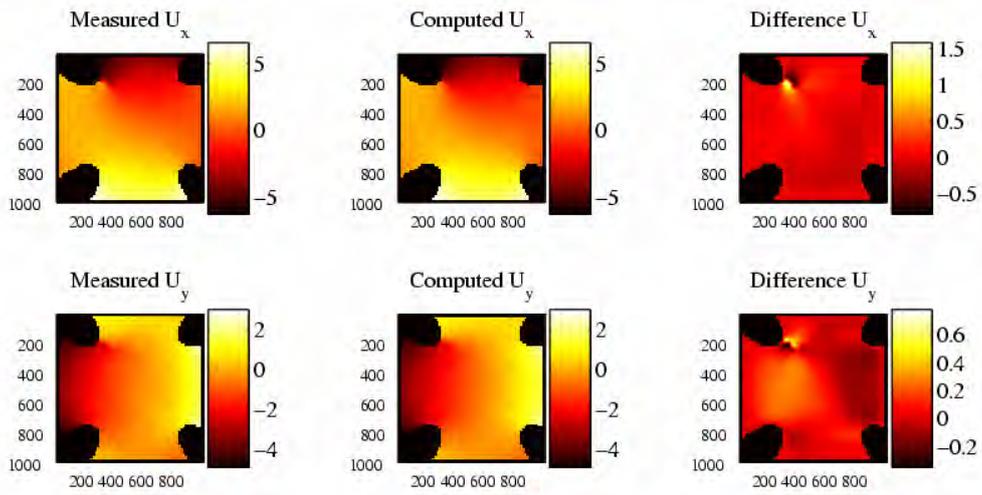

-a-

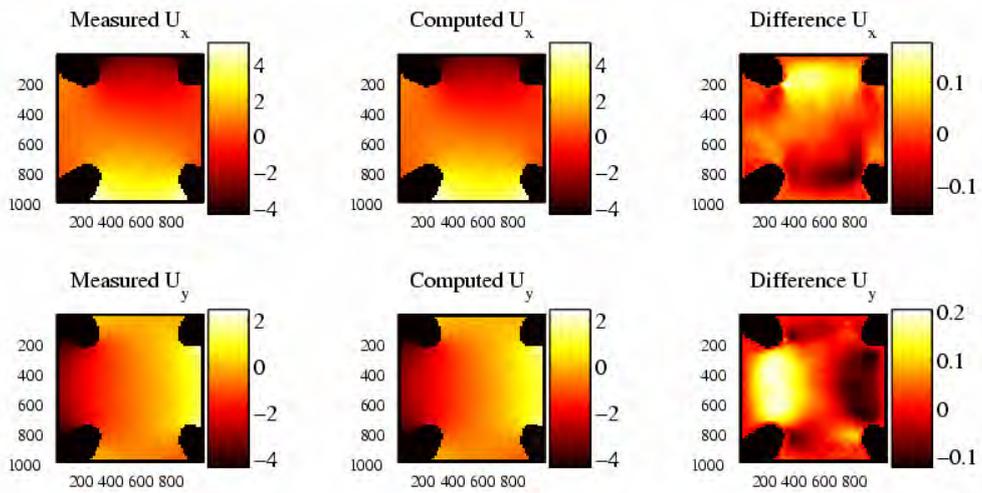

-b-

Fig. 10.

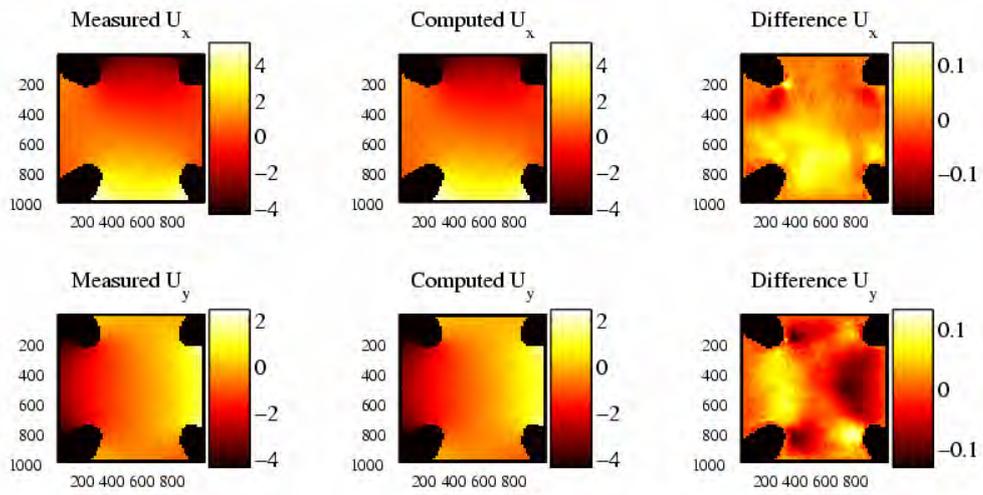

Fig. 11.

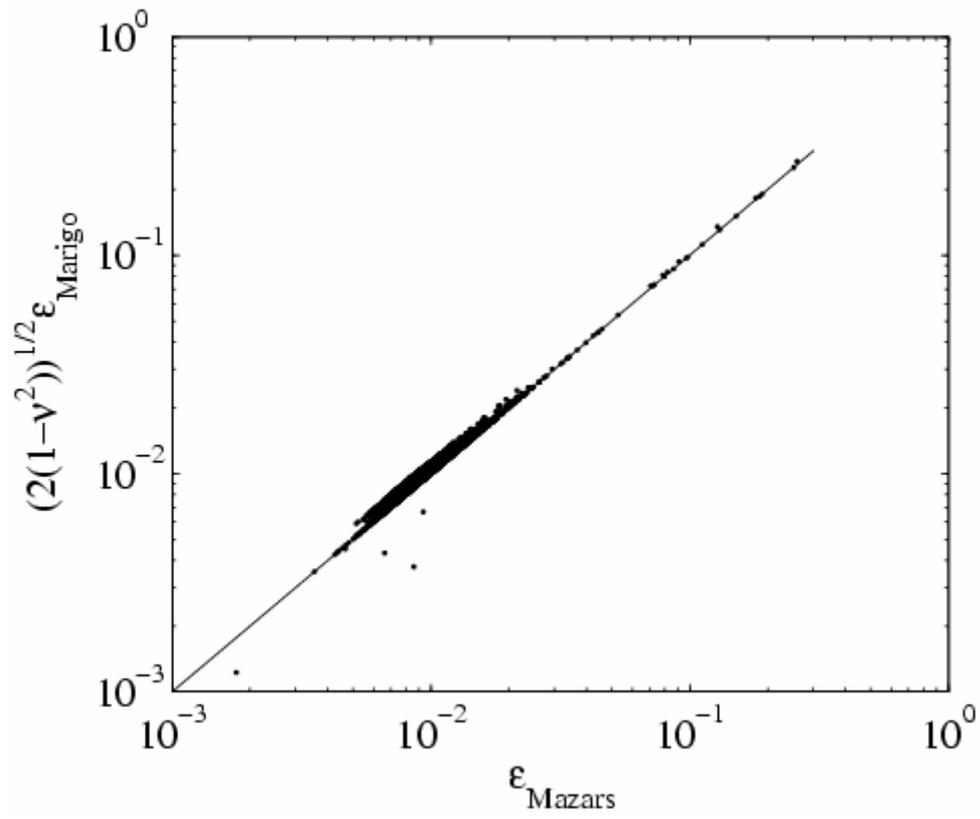

Fig. 12.

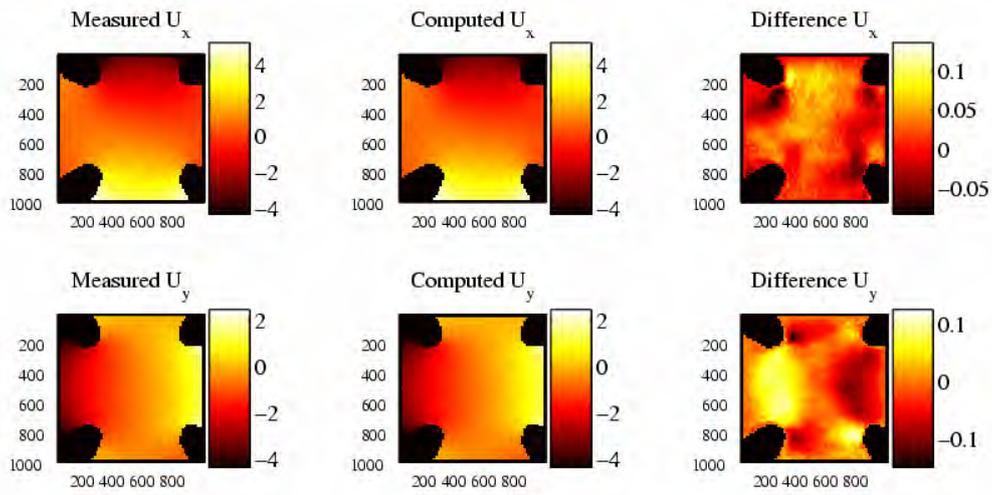

Fig. 13.

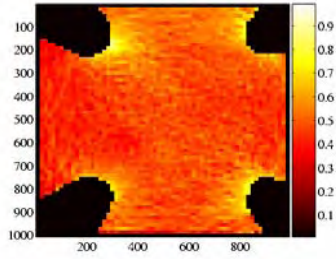 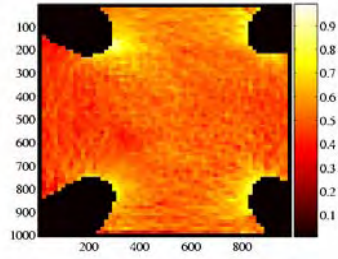 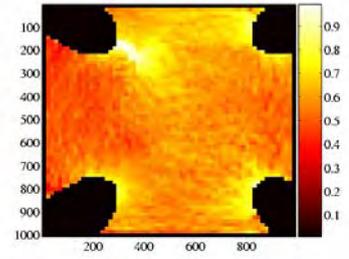

-a-             -b-             -c-

Fig. 14.